\documentclass[prx,10pt,notitlepage,twocolumn]{revtex4}

\usepackage{graphicx}
\usepackage{amsmath}
\usepackage{amssymb}
\usepackage{amsfonts}
\usepackage{color}
\usepackage{xspace}
\usepackage{subfigure}
\usepackage{bm}
\usepackage{bbold}

\newcommand{\ket}[1]{| #1 \rangle}

\begin{document}

\title{Conservation laws, radiative decay rates, and excited state localization in organometallic complexes with strong spin-orbit coupling}

\author{B. J. Powell}\affiliation{Centre for Organic Photonics and Electronics, School of Mathematics and Physics, The University of Queensland, Brisbane, Queensland, 4072, Australia}

\begin{abstract}
There is longstanding fundamental interest in 6-fold coordinated $d^6$ ($t_{2g}^6$) transition metal complexes such as [Ru(bpy)$_3$]$^{2+}$ and Ir(ppy)$_3$, particularly their phosphorescence. This interest has increased with the growing realisation that many of these complexes have potential uses in applications including photovoltaics, imaging, sensing, and light-emitting diodes. In order to design new complexes with properties tailored for specific applications a detailed understanding of the low-energy excited states, particularly the lowest energy triplet state, $T_1$, is required. Here we describe a 
model of pseudo-octahedral complexes based on a pseudo-angular momentum representation and show that the predictions of this model are in excellent agreement with experiment - even when the deviations from octahedral symmetry are large. This model gives a natural explanation of zero-field splitting of $T_1$ and of the relative radiative rates of the three sublevels in terms of the conservation of time-reversal parity and total angular momentum modulo two. We show that the broad parameter regime consistent with the experimental data implies significant localization of the excited state.
\end{abstract}

\maketitle

\section{Introduction}

Six-fold coordinated $d^6$ ($t_{2g}^6$) transition metal complexes, such as those shown in Fig. \ref{fig:sketch}a,b, share many common properties. These include their marked similarities in their low-energy spectra \cite{R1}, cf. Table \ref{tab:expt}, and the competition between localization and delocalizsation in their excited states \cite{R2}. Beyond their intrinsic scientific interest, understanding and controlling this phenomenology is further motivated by the potential for the use of such complexes in diverse applications including dye-sensitized solar cells, non-linear optics, photocatalysis, biological imaging, chemical and biological sensing, photodynamic therapy, light-emitting electro-chemical cells and organic light emitting diodes \cite{R1,R3,R4,R5,R6,R7}. As many of these applications make use of the excited state properties of these complexes a deep understanding of the low-energy excited states, particularly the lowest energy triplet state, $T_1$, is required to enable the rational design of new complexes. 

Coordination complexes where there is strong spin-orbit coupling (SOC) present a particular challenge to theory because of the need to describe both the ligand field and the relativistic effects correctly. There has been significant progress in applying relativistic time-dependent density functional theory (TDDFT) to such complexes; but significant challenges remain, for example correctly describing the zero-field splitting \cite{R8,R9,R10,R11}. There has been less recent focus on the use of semi-empirical approaches, such as ligand field theory \cite{R8,R12,R13,R14}. However, semi-empirical approaches have an important role to play \cite{R15}. Firstly, they provide a general framework to understand experimental and computational results across whole classes of complexes. Secondly, when properly parameterised they can provide accuracy that is competitive with first principles methods. Thirdly, they can provide general design rules that allow one to effectively target new complexes for specific applications.

A long standing question in these complexes is whether the excited state is localized to a single ligand or delocalized  \cite{R2}.
The main semi-empricial approach to understanding organometallic complexes is ligand field theory. Once all of the spatial symmetries are broken there is ligand field theory is limited to a  perturbative regime near approximate symmetries, this makes an accurate description of localised excited states challenging.

In this paper we describe a  semi-empirical approach, based on the pseudo-angular momentum approach that has found widespread use in, e.g., interpreting electron paramagnetic resonance experiments. We derive conservation laws based on the total angular momentum (pseudo plus spin) that apply even when the pseudo-octahedral and trigonal symmetries are strongly broken. These conservation laws imply selection rules for radiative emission. We show that this model reproduces the experimentally measured trends in the radiative decay rates and excitation energies for all of the complexes for which we have data to compare with in the literature. These trends are insensitive to the parameters of the model studied. Finally, we show that for the wide parameter range compatible with experiment the pseudo-angular momentum model predicts significant localization of the excited state.

\section{The pseudo-angular momentum model. }

It has long been understood \cite{R16} that the three-fold degenerate states can be represented by an $l=1$ pseudo-angular momentum. Perhaps the best known example of this are the $t_{2g}$ states of a transition metal in an octahedral ligand field. In the $d^6$ complexes considered here the $t_{2g}$ orbitals are filled, whereas the $e_g$-orbitals are high lying virtual states. Therefore we only include the $t_{2g}$ orbitals in the model described below. 

\begin{figure}
\begin{center}
\includegraphics[width=0.9\columnwidth]{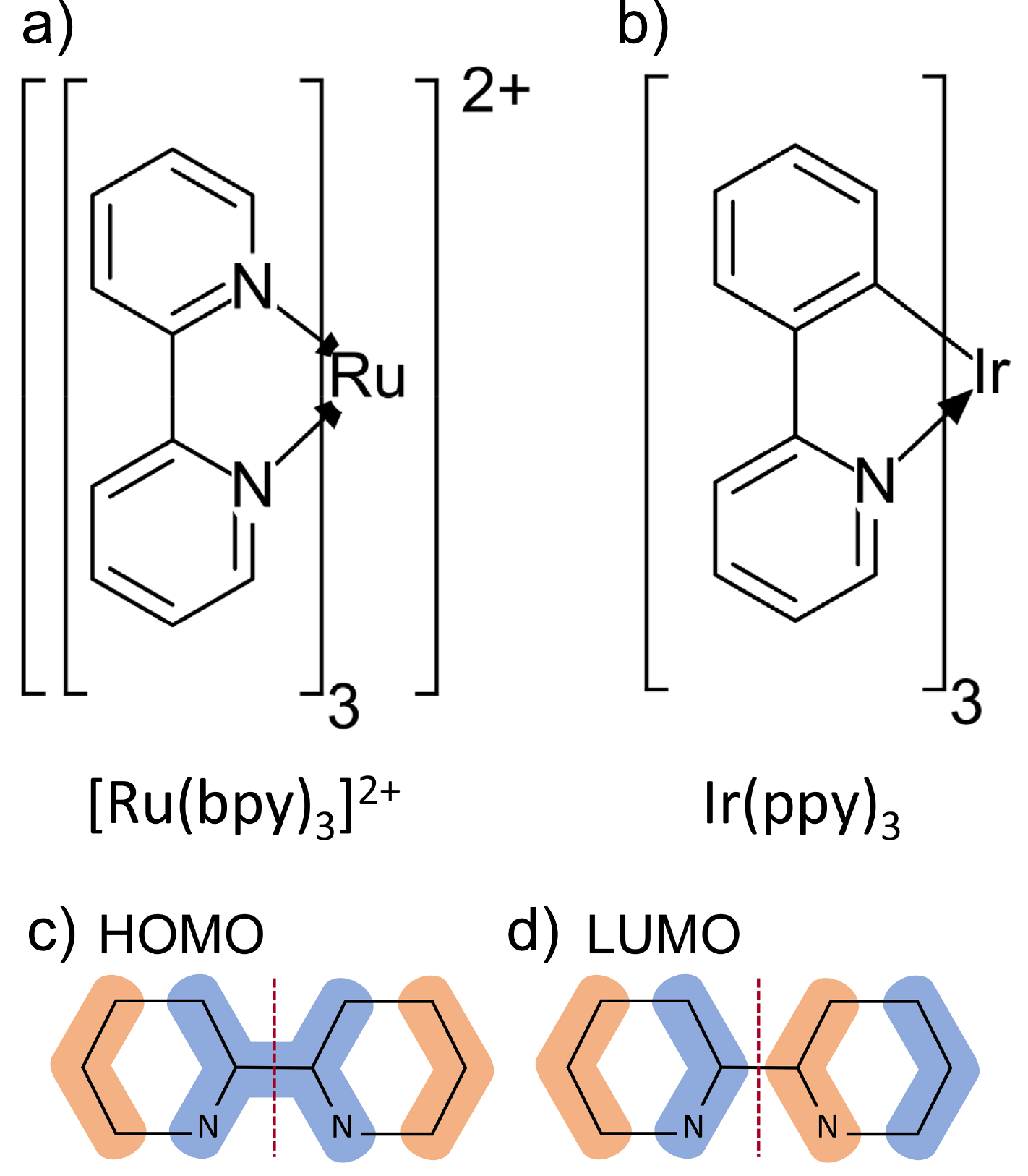}  	
\caption{The structures of two important pseudo-octahedral transition metal complexes: a) [Ru(bpy)$_3$]$^{2+}$ and b) Ir(ppy)$_3$, where bpy is bipyridine and ppy is 2-phenylpyridyl. Sketches of the c) $\pi$ and d) $\pi^*$ orbitals of a bpy ligand with the reflection plane marked by the dashed line. It is clear that these correspond to the bonding and antibonding combinations of singly occupied molecular orbitals of a pyridine radical.  }\label{fig:sketch}
\end{center}
\end{figure}

The complexes listed in Table \ref{tab:expt} have 6-fold coordinated metal atoms, but the ligands break the octahedral symmetry. In complexes with D$_3$ symmetry, e.g., [Ru(bpy)$_3$]$^{2+}$, the ligands have a reflection symmetry, cf. Fig. \ref{fig:sketch}c,d. For a single bpy ligand the highest energy ligand $\pi$-orbitals are even under this reflection whereas the lowest energy $\pi^*$-orbitals are odd under the same reflection, as one would expect from simple symmetry arguments \cite{R13}. Therefore, linear combinations the $\pi$-orbitals transform corresponding to the $t_{2g}$ representation of O$_{\textrm{h}}$, whereas the $\pi^*$-orbitals form a representation of $t_{1u}$. Therefore, $\pi$-orbitals mix effectively with the occupied metal $d$ ($t_{2g}$) orbitals but $\pi^*$-orbitals do not. Thus the highest occupied molecular orbitals (HOMOs), $h_{t_{2g}}^m$, of the complex will have a significant contribution from both the ligand $\pi$-orbitals, $\pi_{t_2g}^m$, and the metal-$t_{2g}$ orbitals, $d_{t_{2g}}^m$. Neglecting smaller contributions from other ligand or metal orbitals, we have
\begin{eqnarray}
h_{t_{2g}}^m\simeq d_{t_{2g}}^m  \cos\theta+\pi_{t_{2g}}^m  \sin\theta,
\end{eqnarray}
where $\theta$ parameterises the degree of mixing, and $m\in\{1,2,3\}$ labels the ligands and symmetry equivalent linear combinations of d-orbitals. In contrast, the lowest unoccupied molecular orbitals (LUMOs) of the complex will be almost pure ligand $\pi^*$-orbitals. 

\begin{table*}
\begin{center}
\begin{tabular}{l|cccccc}
 & $E_{I,II}$ [cm$^{-1}$] & $E_{II,III}$  [cm$^{-1}$] & $\tau_I$ ($\bm{1/k_R^I}$) [$\mu$s] & $\tau_{II}$ ($\bm{1/k_R^{II}}$) [$\mu$s]& $\tau_{III}$ ($\bm{1/k_R^{III}}$) [$\mu$s]\\ \hline
 Ir(biqa)$_3$ & 14 & 64 & 107 ({\bf114}) & 5.6 ({\bf5.7}) & 0.36 ({\bf0.38}) \\ 
Ir(ppy)$_3$ (in PMMA) & 12.2 & 113 & 154 ({\bf175}) & 15 ({\bf17}) & 0.33 ({\bf0.34}) \\ 
Ir(ppy)$_3$ (in CH$_2$Cl$_2$) & 19 & 151 & 116 & 6.4 & 0.2 \\ 
Ir(dm-2-piq)$_2$(acac) & 9.5-10 & 140-150 & 80-124 & 6.5-8.6 & 0.33-0.44 \\ 
$\textrm{[Os(phen)]}_2\textrm{(dppm)]}^{2+}$  & 16 & 106 & 95 & 13 & 0.6 \\ 
$\textrm{[Os(phen)}_2\textrm{(dpae)]}^{2+}$ & 21 & 92 & 100 & 10 & 0.7 \\ 
Ir(piq)(ppy)$_2$ & 16 & 91 & 64 & 10.5 & 0.3 \\ 
Ir(4,6-dFppy)$_2$(acac) & 16 & 93 & 44 & 9 & 0.4 \\ 
Ir(pbt)$_2$(acac) & 6 & 97 & 82 & 25 & 0.4 \\ 
Ir(piq)$_2$(acac) & 9 & 87 & 47 & 8 & 0.3 \\ 
$\textrm{[Os(dpphen)}_2\textrm{(dpae)]}^{2+}$ & 19 & 75 & 92 & 9 & 0.7 \\ 
$\textrm{[Os(phen)}_2\textrm{(DPEphos)]}^{2+}$ & 16 & 68 & 104 & 14 & 0.9 \\ 
$\textrm{[Os(phen)}_2\textrm{(dppe)]}^{2+}$ & 19 & 55 & 107 & 12 & 0.9 \\ 
Ir(piq)$_2$(ppy) & 9 & 56 & 60 & 6.4 & 0.44 \\ 
$\textrm{[Os(phen)}_2\textrm{(dppene)]}^{2+}$ & 18 & 46 & 108 & 15 & 1.1 \\ 
$\textrm{[Ru(ppy)}_3\textrm{]}^{2+}$ & 8.7 & 52 & 230 & 8 & 0.9 \\ 
Ir(piq)$_3$ & 11 & 53 & 57 & 5.3 & 0.42 \\ 
Ir(4,6-dFppy)$_2$(pic) & 9 & 67 & 47 & 21 & 0.3 \\ 
Ir(thpy)$_2$(acac) & 3.5 & 31 & 113 & 35 & 1.5 \\ 
Ir(ppy)$_2$(ppy-NPH$_2$) & 6 & 21 & 188 & 19 & 1.8 \\ 
Ir(ppy-NPH$_2$)$_3$ & 6 & 20 & 177 & 15 & 1.4 \\ 
Ir(ppy)(ppy-NPH$_2$)$_2$ & 6 & 17 & 163 & 20 & 2 \\ 
Ir(btp)$_2$(acac) & 2.9 & 22 & 150 & 58 & 2 \\ 
Ir(btp)$_2$(acac) & 2.9 & 11.9 & 62 & 19 & 3 \\ 
Ir(s1-thpy)$_2$(acac) & 3 & 13 & 128 & 62 & 3 \\ 
Ir(ppy)$_2$(CO)(Cl) & $<1$ & $<1$ & 300 & 85 & 9 \\ 
$\textrm{[Rh(ppy)}_3\textrm{]}^{3+}$ & - & - & $4.5\times10^3$ & $1.35\times10^3$ & 650
\end{tabular}
\end{center}
\caption{Key spectroscopic data for pseudo-octahedral $d^6$-complexes. $E_{I,II}$ is the energy gap between the two lowest energy substates of $T_1$, $E_{II,III}$ is the energy gap between the two highest energy substates of $T_1$ and the total lifetime of substate $m$ $\tau_m=(k_R^m+k_{NR}^m )^{-1}$, where $k_R^m$ and $k_{NR}^m$ and the radiative and non-radiative lifetimes of substate $m$. For Ir(ppy)$_3$ and Ir(biqa)$_3$ we also list $1/k_R^m$ (in bold) which, unsurprisingly given the high photoluminescent quantum yields in these complexes, shows the same trend as $\tau_m$. We are not aware of measurements of $k_R^m$ in other relevant complexes. Note that in all complexes $E_{I,II}<E_{II,III}$ and $\tau_I>\tau_{II}>\tau_{III}$, which suggests that $k_R^I<k_R^{II}<k_R^{III}$. To avoid selection bias we have included all and only those pseudo-octahedral $d^6$-complexes included in Table 2 of the recent review by Yersin et al. \cite{R1}. The two rows for Ir(btp)$_2$(acac) correspond to different sites.}\label{tab:expt}
\end{table*}

Low energy excited states can be well approximated by a single hole in the HOMO manifold and a single electron in the LUMO manifold \cite{R17}. As both the HOMOs and LUMOs of the complex are three-fold degenerate one can label such states by two $l=1$ pseudo-angular momenta, which we denote $\bm{L}_H$ and $\bm{L}_L$ respectively.  We will  only discuss this assignment for three real space HOMO spin orbitals, $h_{t_2g}^m$ -- it is trivial to extend the following analysis to the LUMOs. By referring to these states as `HOMOs' and `LUMOs' we are adopting the language of molecular orbital theory. However, we note that so-long as the $h_{t_{2g}}^m$ are three local states related by rotations of $2\pi/3$ the discussion below goes through regardless of the degree of correlations in the states. It is therefore convenient to work in second quantised notation, so we define $\langle \bm{r}|a_m^\dagger|0\rangle=h_{t_{2g}}^m$, where $|0\rangle$ is the ground state, $|\bm{r}\rangle$ is the state with a hole at position $\bm{r}$; spin labels are supressed.

We introduce three `Bloch' operators defined by
\begin{eqnarray}
b_k^+=\textrm{sgn}^k(-k)\frac{1}{\sqrt{3}} \sum_m a_m^\dagger e^{i2\pi km/3},
\end{eqnarray}
where $k\in\{-1,0,1\}$. Finally we identify the states created by the Bloch operators with the eigenstates of $L_H^z$, i.e., $\langle0|b_k L^z_H b_k^\dagger|0\rangle=k$. The phase pre-factors [$\textrm{sgn}^k(-k)$]  in the definition of the Bloch operators are required to allow this assignment and maintain the required behaviour under time reversal symmetry. 

As the LUMOs are pure ligand orbitals the exchange interaction will be dominated by the exchange interaction between the ligand $\pi$ and $\pi^*$-orbitals, $J_\pi$. In contrast the SOC on the metal, $\lambda_d$, is much stronger than the SOC on the ligands. Therefore states with one hole in the HOMO and one electron in the LUMO are described by the Hamiltonian
\begin{eqnarray}
H_o=J\bm{S}_H\cdot\bm{S}_L+\lambda\bm{L}_H\cdot\bm{S}_H,
\end{eqnarray}
where $\bm{S}_H$ is the (net) spin of the electrons in the HOMO, $\bm{S}_L$, is the spin of the electron in the LUMO, $J\simeq J_\pi\sin^2\theta$ and $\lambda\simeq\lambda_d\cos^2\theta$. Thus we expect positive $J$ and $\lambda$. 

If the excited state is sufficiently long lived for the geometry to relax it will be unstable to a Jahn-Teller distortion, which lifts the degeneracy. In terms of the pseudo-angular momenta this can be represented via the terms
\begin{eqnarray}
H_{JT}&=&\delta Q\left[(L_H^x )^2-(L_H^y )^2\right]+\gamma Q\left[(L_L^x )^2-(L_L^y )^2\right]\notag\\&&+kQ^2,
\end{eqnarray}
where $Q$ is the coordinate of the rhombic distortion perpendicular to the C$_3$-axis of the complex, $\delta$ ($\gamma$) is the coupling constant to the HOMOs (LUMOs) and $k$ is the spring constant of the Jahn-Teller mode. 

It is helpful to briefly discuss the Jahn-Teller effect in the pseudo-angular momentum language, as this is not entirely intuitive. Consider the term 
\begin{eqnarray}
[(L_H^x )^2-(L_H^y )^2 ]=\frac12 [(L_H^+ )^2+(L_H^- )^2 ]
\end{eqnarray}
In terms of the Bloch operators $L_H^\dagger=2(b_1^\dagger b_0+b_0^\dagger b_{-1})$ and $L_H^-=2(b_{-1}^\dagger b_0+b_0^\dagger b_1 )$. Hence, 
\begin{widetext}
\begin{eqnarray}
(L_H^x )^2-(L_H^y )^2&=&b_1^\dagger b_{-1}+b_{-1}^\dagger b_{-1}\notag\\
&=&-\frac13 \Big[2a_1^\dagger a_1-a_2^\dagger a_2-a_3^\dagger a_3  
+2(a_2^\dagger a_3+a_3^\dagger a_2 )-a_1^\dagger a_2  
-a_2^\dagger a_1-a_1^\dagger a_3-a_3^\dagger a_1 \Big].
\end{eqnarray}
\end{widetext}
It is therefore clear that this physics of $H_{JT}$ is that of the $T\times t$ Jahn-Teller problem [or, once trigonal terms are included, below, the $(A+E)\times e$ pseudo-Jahn-Teller problem] and that this distortion corresponds to the so-called $E_\theta$ distortion in the notation of, e.g., section 5 of Ref.  \cite{R18}. The $E_\varepsilon$ distortion corresponds to terms proportional to $(1/2)[(L_H^+)^2-(L_H^-)^2 ]=i\{L_H^x,L_H^y \}$, where curly brackets indicate anticommutation.

 In general the Jahn-Teller distortion could also induce a trigonal component of the distortion [which would couple to $(L_\nu^z )^2$, where $\nu=H$ or $L$], however this does not produce any qualitatively new features and so, for simplicity, we neglect it below. Thence, the form of $H_{JT}$ is constrained to the form given above by symmetry as: (1) terms that are proportional to odd powers of $L_\nu^\beta$, where $\beta=x,\,y$ or $z$ break time reversal symmetry and so may not appear in the Hamiltonian for scalar $Q$ and (2) for $l=1$ any even power of $L_\nu^\beta$ is proportional to $(L_\nu^\beta)^2$.

However, the complexes in Table \ref{tab:expt} are not octahedral, but trigonal. In terms of the pseudo-angular momenta, this introduces the additional terms
\begin{eqnarray}
H_t=\Delta(L_H^z )^2+\Gamma(L_L^z )^2,
\end{eqnarray}
where $\Delta$ ($\Gamma$) is the energy differences between the HOMO and HOMO-1 (LUMO and LUMO+1) in the trigonal ground state, $S_0$, geometry. Indeed, it immediately follows from time reversal symmetry that the trigonal terms in the Hamiltonian are constrained to take this form. $t_{2g}\rightarrow a_1+e$ and $t_{1u}\rightarrow a_2+e$ on lowering the symmetry from O$_{\textrm{h}}$ to D$_3$. Therefore, the two pairs of $e$ states are allowed to weakly mix, stabilising the $L_H^z=\pm1$ states and destabilising the $L_L^z=\pm1$ states. Thus one expects that both $\Delta$ and $\Gamma$ will be positive \cite{R13}. The approximate D$_3$ symmetry of the complexes with lower symmetry, e.g. C$_3$, complexes considered here means that we expect both parameters to remain positive for all of the complexes considered here \cite{footKM}.
Thus the effective pseudo-angular momentum Hamiltonian for the low-energy excitations is 
\begin{eqnarray}
H=H_o+H_t+H_{JT}.
\end{eqnarray}

By definition $Q=0$ in the $S_0$ geometry and, by suitably rescaling the parameters, one may define $Q=1$ in the $T_1$ geometry. Similarly any trigonal component to the Jahn-Teller distortion can be taken simply to shift the value of $\Delta$ ($\Gamma$).  Therefore, up to constants, in the $T_1$ geometry the effective electronic Hamiltonian is
\begin{eqnarray}
H&=&J\bm{S}_H\cdot\bm{S}_L+\lambda\bm{L}_H\cdot\bm{S}_H+\Delta(L_H^z )^2+\Gamma(L_L^z )^2\notag\\&&+\delta[(L_H^x )^2-(L_H^y )^2 ]+\gamma[(L_L^x )^2-(L_L^y )^2 ].\label{eqn:H}
\end{eqnarray}
As well as describing systems displaying a Jahn-Teller distortion, this model is also appropriate for heteroleptic complexes. Indeed for appropriate choices of $\Delta$, $\delta$, $\Gamma$ and $\gamma$ one can parameterise arbitrary energy differences of the frontier orbitals. We discuss the values of these parameters in the Appendix. On the basis of this discussion, for Ir(ppy)$_3$, we take $\lambda/J=0.2$ and $\Delta/J=0.5$, with $J\sim1$~eV; $\delta\lesssim\Delta$ and $\gamma\lesssim\Gamma$ below. Clearly, for example, $\lambda$ is strongly dependent on the transition metal in question. However, our main qualitative results are insensitive to the values of these parameters -- to emphasize this we explore a wide range of other parameters in the sup. info.

\section{RESULTS}

\subsection{Octahedral model} 

Before considering the full pseudo-angular momentum model, $H$, it is important to understand the symmetries of $H_o$. (i) $\bm{L}_L$ does not couple to any of the other variables. Therefore, $\bm{L}_L^2$ and $L_L^z$ are good quantum numbers. (ii) We can define a `total' angular momentum, $\bm{I}=\bm{L}_H+\bm{S}$, where $\bm{S}=\bm{S}_H+\bm{S}_L$. $I^2$ and $I^z$ commute with $H_o$ therefore $I$ and $I^z$ are also good quantum numbers. 

\begin{figure}
\begin{center}
\includegraphics[width=0.9\columnwidth]{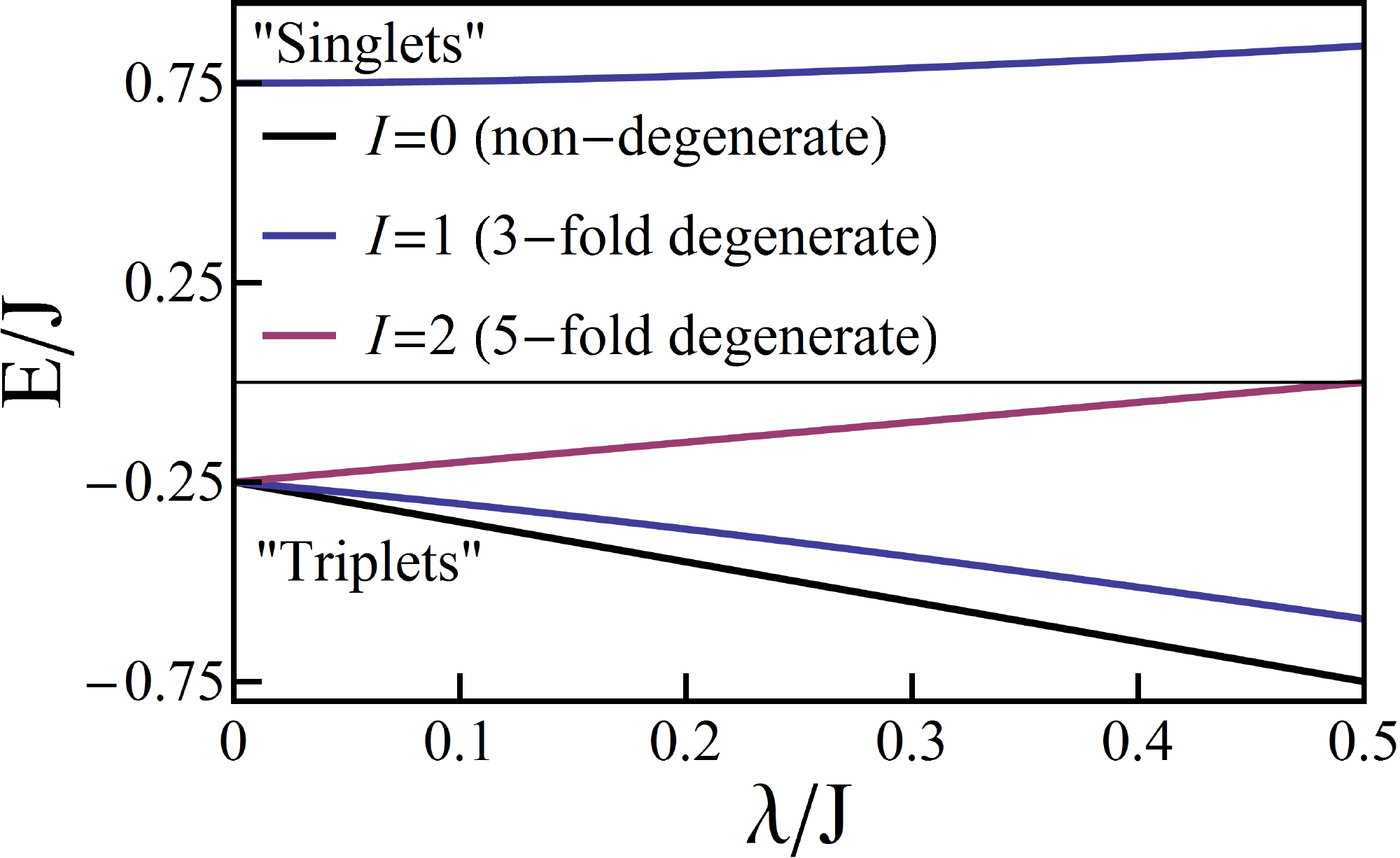}  	
\caption{Energy eigenvalues of $H_o$ for states with $L_z=0$. At $\lambda=0$ the singlets have $E=3J/4$ and the triplets have $E=-J/4$. For $\lambda>0$ the labels ``singlet" and ``triplet" are no longer strictly defined (in their usual sense) nevertheless the relatively small energy shifts suggest that these labels retain some meaning, this claim is supported by directly examining the character of the eigenstates. It is interesting to note that, already in the octahedral problem, the lowest energy (non-degenerate) state has no singlet contribution to its wavefunction for any value of $\lambda$, thus radiative transitions from this state are forbidden. }\label{fig:oct}
\end{center}
\end{figure}

We plot the energies of the exact solutions of $H_o$ in Fig. \ref{fig:oct} (Table \ref{Tab:basis} gives the basis used for all calculations in the paper). For simplicity Fig. \ref{fig:oct} shows only the solutions with $L_L^z=0$ -- because $\bm{L}_L$ is decoupled from the other angular momenta it can be immediately seen that the other solutions simply triple the degeneracies of all states.  Note that, firstly, the spectrum of $H_o$ is not very similar to those of the pseudo-octahedral complexes we are seeking to model. However, this model is an important stepping stone to understanding the full Hamiltonian. Secondly, the eigenstates can be classified by their total angular momentum quantum number, $I$, and, as $H_o$ is SU(2) symmetric, have the expected $2I+1$ degeneracy. Thirdly, all of the singlets have $I=1$; as $L_H=1$ and, by definition, singlets have $S=0$. This means that, regardless of how strong the SOC is, the singlets can only mix with the $I=1$ triplets. Therefore radiative decay from the $I=0$ and $I=2$ triplets is forbidden by the conservation of $I$.

\begin{table*}
\begin{center}
\begin{tabular}{l|c|c|c|c|c}
Name & \begin{tabular}{c}
Relationship to eigenstate\\ of $H_o$ when $\lambda=0$,\\ $\ket{I,I^z,S}$ 
\end{tabular}
  & $\cal T$ & ${\cal I}^z$ & $\ket{{S}_L^z{S}_H^z{L}_H^z}$ & \begin{tabular}{c}Singlets \\mixed with \\in full model \end{tabular}
\\  \hline
$\ket{S_z}$ & $\ket{1,0,0}$ & -1 & 1 & $\frac{1}{\sqrt{2}}\left(\ket{\uparrow\downarrow\Rightarrow}-\ket{\downarrow\uparrow\Rightarrow}\right)$ & - 
\\
$\ket{S_x}$ & $\frac{1}{\sqrt{2}}\left(\ket{1,1,0}-\ket{1,-1,0}\right)$ & 1 & -1 & $\frac{1}{2}\left(\ket{\uparrow\downarrow\Uparrow}-\ket{\downarrow\uparrow\Uparrow}-\ket{\uparrow\downarrow\Downarrow}+\ket{\downarrow\uparrow\Downarrow}\right)$ & - 
\\
$\ket{S_y}$ & $\frac{1}{\sqrt{2}}\left(\ket{1,1,0}+\ket{1,-1,0}\right)$ & -1 & -1 & $\frac{1}{2}\left(\ket{\uparrow\downarrow\Uparrow}-\ket{\downarrow\uparrow\Uparrow}+\ket{\uparrow\downarrow\Downarrow}-\ket{\downarrow\uparrow\Downarrow}\right)$ & - 
\\ \hline
$\ket{T_\mathbb{1}}$ & $\ket{0,0,1}$ & 1 & 1 & $\frac{1}{\sqrt{3}}\left(\ket{\uparrow\uparrow\Downarrow}+\ket{\downarrow\downarrow\Uparrow}\right)-\frac{1}{\sqrt{6}}\left(\ket{\uparrow\downarrow\Rightarrow}+\ket{\downarrow\uparrow\Rightarrow}\right)$ & None 
\\\hline
$\ket{T_z}$ & $\ket{1,0,1}$ & -1 & 1 & $\frac{1}{\sqrt{2}}\left(\ket{\uparrow\uparrow\Downarrow}-\ket{\downarrow\downarrow\Uparrow}\right)$ & $\ket{S_z}$ 
\\
$\ket{T_x}$ & $\frac{1}{\sqrt{2}}\left(\ket{1,1,1}-\ket{1,-1,1}\right)$ & -1 & -1 & $\frac{1}{2}\left[\ket{\uparrow\uparrow\Rightarrow}-\ket{\downarrow\downarrow\Rightarrow}-\frac{1}{\sqrt{2}}\left(\ket{\uparrow\downarrow\Uparrow}+\ket{\downarrow\uparrow\Uparrow}-\ket{\uparrow\downarrow\Downarrow}-\ket{\downarrow\uparrow\Downarrow}\right)\right]$ & $\ket{S_y}$ 
\vspace*{2pt}\\
$\ket{T_y}$ & $\frac{1}{\sqrt{2}}\left(\ket{1,1,1}+\ket{1,-1,1}\right)$ & 1 & -1 & $\frac{1}{2}\left[\ket{\uparrow\uparrow\Rightarrow}+\ket{\downarrow\downarrow\Rightarrow}-\frac{1}{\sqrt{2}}\left(\ket{\uparrow\downarrow\Uparrow}+\ket{\downarrow\uparrow\Uparrow}+\ket{\uparrow\downarrow\Downarrow}+\ket{\downarrow\uparrow\Downarrow}\right)\right]$ & $\ket{S_x}$ 
\vspace*{1pt}\\ \hline
$\ket{T_{z^2}}$ & $\ket{2,0,1}$ & 1 & 1 & $\frac{1}{\sqrt{6}}\left(\ket{\uparrow\uparrow\Downarrow}+\ket{\downarrow\downarrow\Uparrow}\right)+\frac{1}{\sqrt{3}}\left(\ket{\uparrow\downarrow\Rightarrow}+\ket{\downarrow\uparrow\Rightarrow}\right)$ & None
\\
$\ket{T_{xz}}$ & $\frac{1}{\sqrt{2}}\left(\ket{2,1,1}-\ket{2,-1,1}\right)$ & -1 & -1 & $\frac{1}{2}\left[\ket{\uparrow\uparrow\Rightarrow}-\ket{\downarrow\downarrow\Rightarrow}+\frac{1}{\sqrt{2}}\left(\ket{\uparrow\downarrow\Uparrow}+\ket{\downarrow\uparrow\Uparrow}-\ket{\uparrow\downarrow\Downarrow}-\ket{\downarrow\uparrow\Downarrow}\right)\right]$ & $\ket{S_y}$ 
\vspace*{2pt}\\
$\ket{T_{yz}}$ & $\frac{1}{\sqrt{2}}\left(\ket{2,1,1}+\ket{2,-1,1}\right)$ & 1 & -1 & $\frac{1}{2}\left[\ket{\uparrow\uparrow\Rightarrow}+\ket{\downarrow\downarrow\Rightarrow}+\frac{1}{\sqrt{2}}\left(\ket{\uparrow\downarrow\Uparrow}+\ket{\downarrow\uparrow\Uparrow}+\ket{\uparrow\downarrow\Downarrow}+\ket{\downarrow\uparrow\Downarrow}\right)\right]$ & $\ket{S_x}$ 
\\
$\ket{T_{xy}}$ & $\frac{1}{\sqrt{2}}\left(\ket{2,2,1}-\ket{2,-2,1}\right)$ & -1 & 1 & $\frac{1}{\sqrt{2}}\left(\ket{\uparrow\uparrow\Uparrow}-\ket{\downarrow\downarrow\Downarrow}\right)$ & $\ket{S_z}$ 
\\
$\ket{T_{x^2-y^2}}$ & $\frac{1}{\sqrt{2}}\left(\ket{2,2,1}+\ket{2,-2,1}\right)$ & 1 & 1 & $\frac{1}{\sqrt{2}}\left(\ket{\uparrow\uparrow\Uparrow}+\ket{\downarrow\downarrow\Downarrow}\right)$ & None 
\end{tabular}
\end{center}
\caption{The basis set used in this paper. The wavefunctions are given in the form $\ket{{S}_L^z{S}_H^z{L}_H^z}$ with $\uparrow$ ($\downarrow$) indicating $S_\nu^z=+1/2$ $(-1/2)$ and $\Uparrow$, $\Rightarrow$ and $\Downarrow$ indicating $L_H^z=1,0$ and $-1$ respectively. ${\cal I}^z=(-1)^{I^z}$. In this table we list only the $L_L^z=0$ [${\cal L}^z=(-1)^{L_L^z}=1$] states. Each state has two partners with $L_L^z=1$ and hence ${\cal L}^z=-1$. The latter two, but not the former, mix under the action of the full Hamiltonian.}\label{Tab:basis}
\end{table*}%

\subsection{Trigonal model} 

\begin{figure}
\begin{center}
\includegraphics[width=0.9\columnwidth]{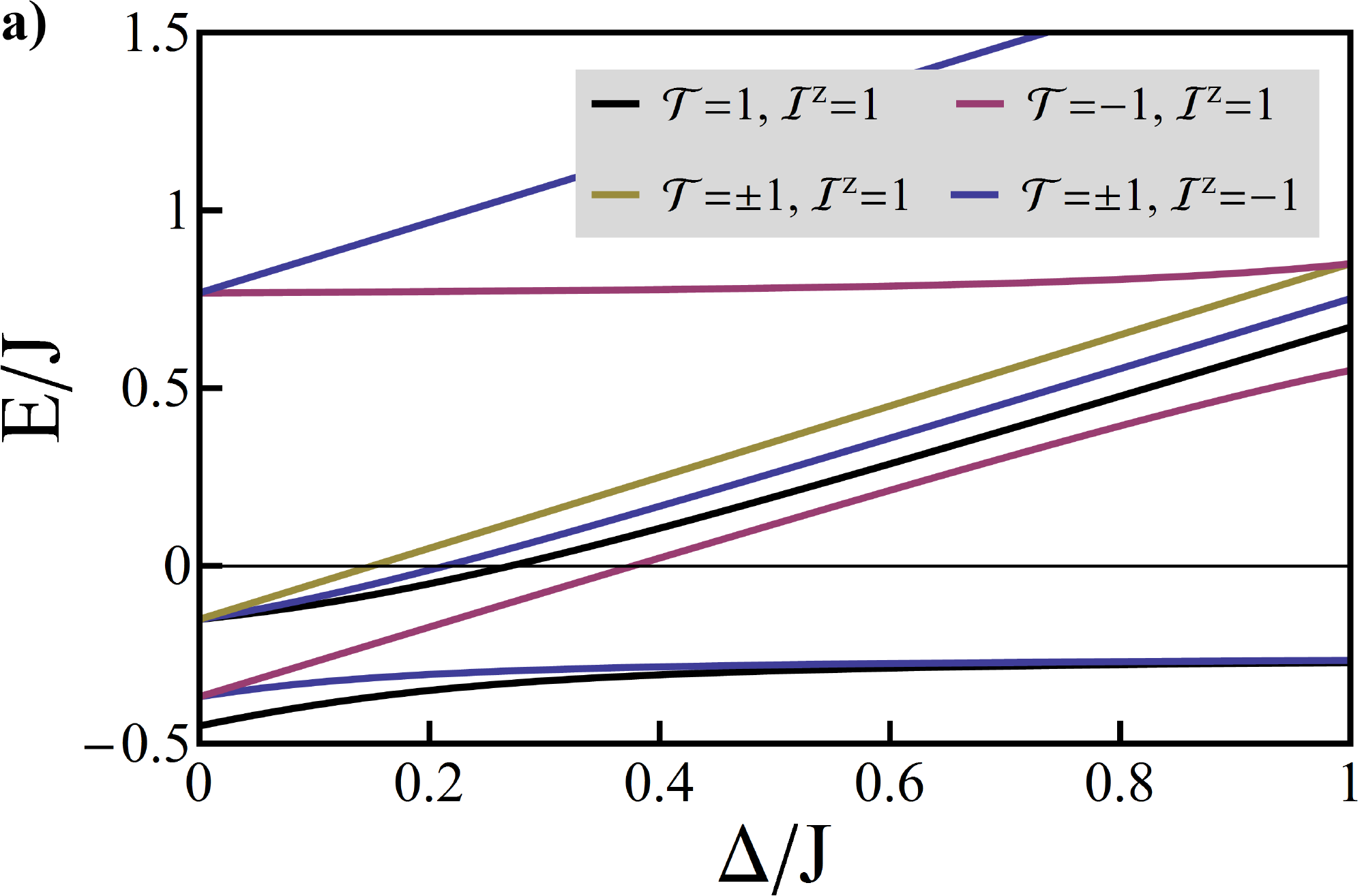}  \vspace*{10pt}\\	
\includegraphics[width=0.9\columnwidth]{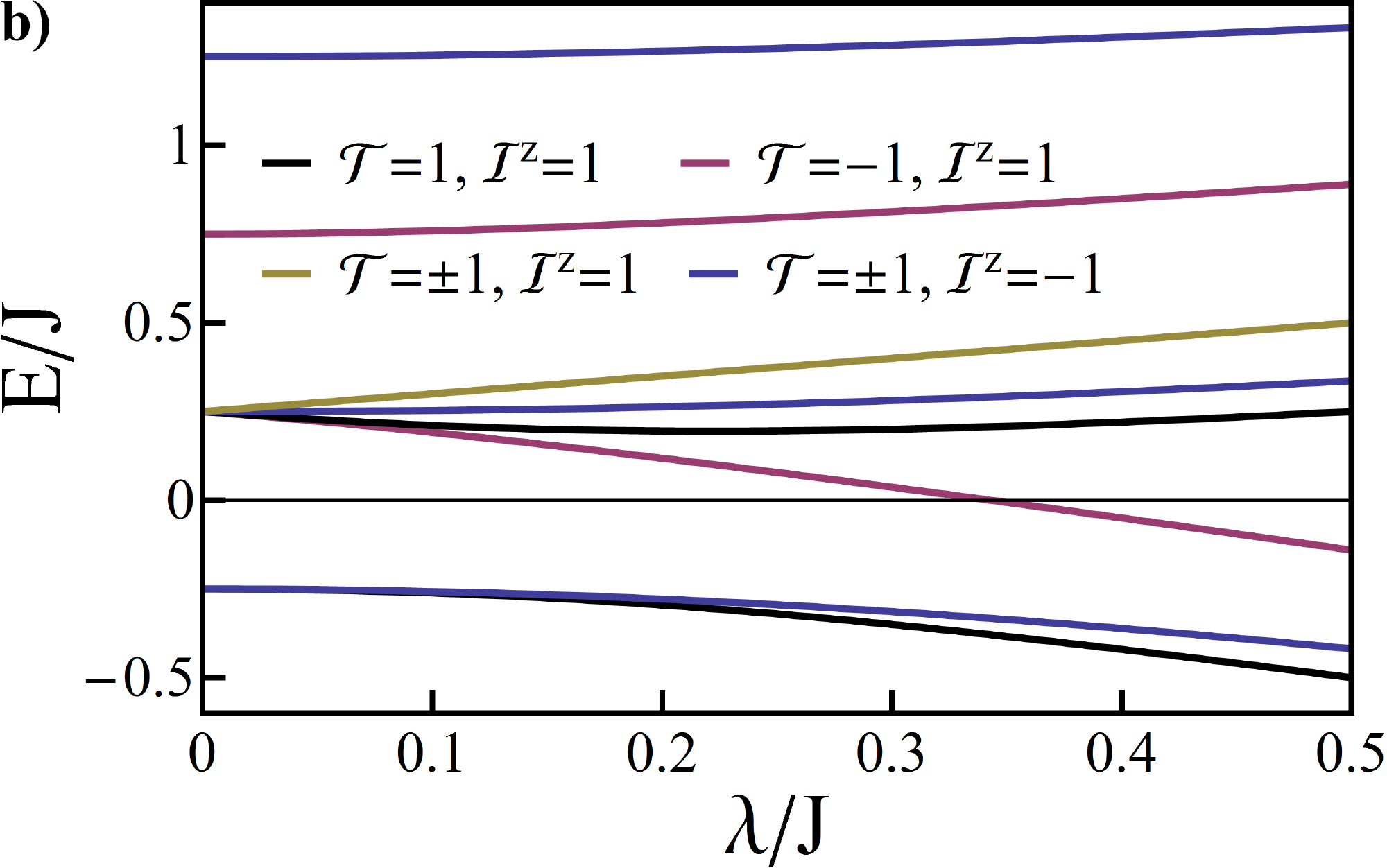}  	
\caption{Solution of the pseudo-angular momentum model of a trigonal complex. a) spectra for $\lambda=J/5$ and varying $\Delta/J$; b) spectra for $\Delta=J/2$ and varying $\lambda/J$. The quantum numbers of the states are also indicated. In both panels the states with quantum numbers labelled as ${\cal T}=\pm1$ are two-fold degenerate. The eigenstates with $L_z=\pm1$ (not shown for clarity) have the same properties except that their energies are increased by $\Gamma$ and all of the degeneracies are doubled corresponding to the two values of $L_z=\pm1$.}\label{fig:trig}
\end{center}
\end{figure}

In Fig. \ref{fig:trig} we plot the spectrum of the trigonal model with no Jahn-Teller distortion, $H_o+H_t$. Again, for simplicity, we only show the solutions with $L_L^z=0$. In this case each state has partners with $L_L^z=\pm1$ that have energies that are higher by $\Gamma$ and display twice the degeneracy of the $L_L^z=0$ state. The trigonal terms break the SU(2) symmetry of the octahedral model and therefore lift the three- and five-fold degeneracies. The calculated spectra are now like those calculated from first-principles for relevant complexes. For example, if trigonal symmetry is enforced for, e.g., [Os(bpy)$_3$]$^{2+}$, Ir(ppy)$_3$, Ir(ptz)$_3$ relativistic TDDFT calculations predict that SOC splits $T_1$ into a non-degenerate state (I) and, at slightly higher energies, a pair of degenerate states (II and III) \cite{R8,R9,R11}.

We saw above that in the octahedral model radiative decay from the lowest energy excited state (I$\rightarrow$0) is forbidden by the conservation of $I$. Because $H_t$ breaks the SU(2) symmetry of the octahedral model $I^2$ no longer commutes with $H$, nevertheless $I^z$ and $L_L^z$ remains a good quantum numbers for the trigonal model. Furthermore, the Hamiltonian is time reversal symmetric, therefore the parity of an eigenstate under time reversal, ${\cal T}=\pm1$, is also a good quantum number. Note however, that $I^z$ does not commute with time reversal so it is not, in general, possible to form states that are simultaneously eigenstates of both. However, one may define states that are simultaneous eigenstates of the ${\cal T}$ and ${\cal I}^z=(-1)^{I^z}$. Therefore, we take these as our quantum numbers, cf. Table \ref{Tab:basis}.

For all parameters studied substate I is composed of the basis state $|T_{\mathbb{1}}\rangle$  admixed with $|T_{z^2}\rangle$ and has quantum numbers ${\cal I}^z={\cal T}=+1$, $L_L^z=0$ whereas states II and III are a degenerate pair with ${\cal I}^z=-1$, ${\cal T}=\pm1$, $L_L^z=0$, cf. Fig. \ref{fig:trig}, whose largest contributions come from $|T_x\rangle$ and $|T_y\rangle$. The singlet states with the same quantum numbers contribute to substates II and III, but all of the singlets are forbidden from mixing with substate I by the combination of time reversal symmetry and the conservation of ${\cal I}^z$. Hence, the I$\rightarrow$0 transition remains forbidden in the trigonal model. Both experiments \cite{R1,R19} and relativistic TDDFT calculations \cite{R8,R9,R10,R11} find that the radiative rates for the transitions II$\rightarrow$0 and III$\rightarrow$0 are more than an order of magnitude faster than that for I$\rightarrow$0, cf. Table \ref{Tab:basis}. The small non-zero decay rate for I$\rightarrow$0 may arise from either Herzberg-Teller coupling \cite{R1,R19} or mixing of state I with higher energy singlet states, which are not included in the pseudo-angular momentum model \cite{R8,R9,R10,R11}.

\subsection{Full model}

Finally, we turn to the full pseudo-angular momentum model, $H$ [Eq. (\ref{eqn:H})]. $I^z$ does not commute with $H_{JT}$. However, $(L_H^x )^2-(L_H^y )^2=\frac12[(L_H^+ )^2+(L_H^- )^2 ]$, where the ladder operators are given by $L_H^\pm=L_H^x\pm iL_H^y$, therefore $I^z$ is conserved modulo two. Thus, ${\cal I}^z$ is conserved even for a trigonal system that has undergone a Jahn-Teller distortion, cf. Table \ref{Tab:basis}. Similarly $L_L^z$ is conserved modulo two, which gives rise to the quantum number ${\cal L}^z=(-1)^{L_L^z}$.

\begin{figure}
\begin{center}
\includegraphics[width=0.9\columnwidth]{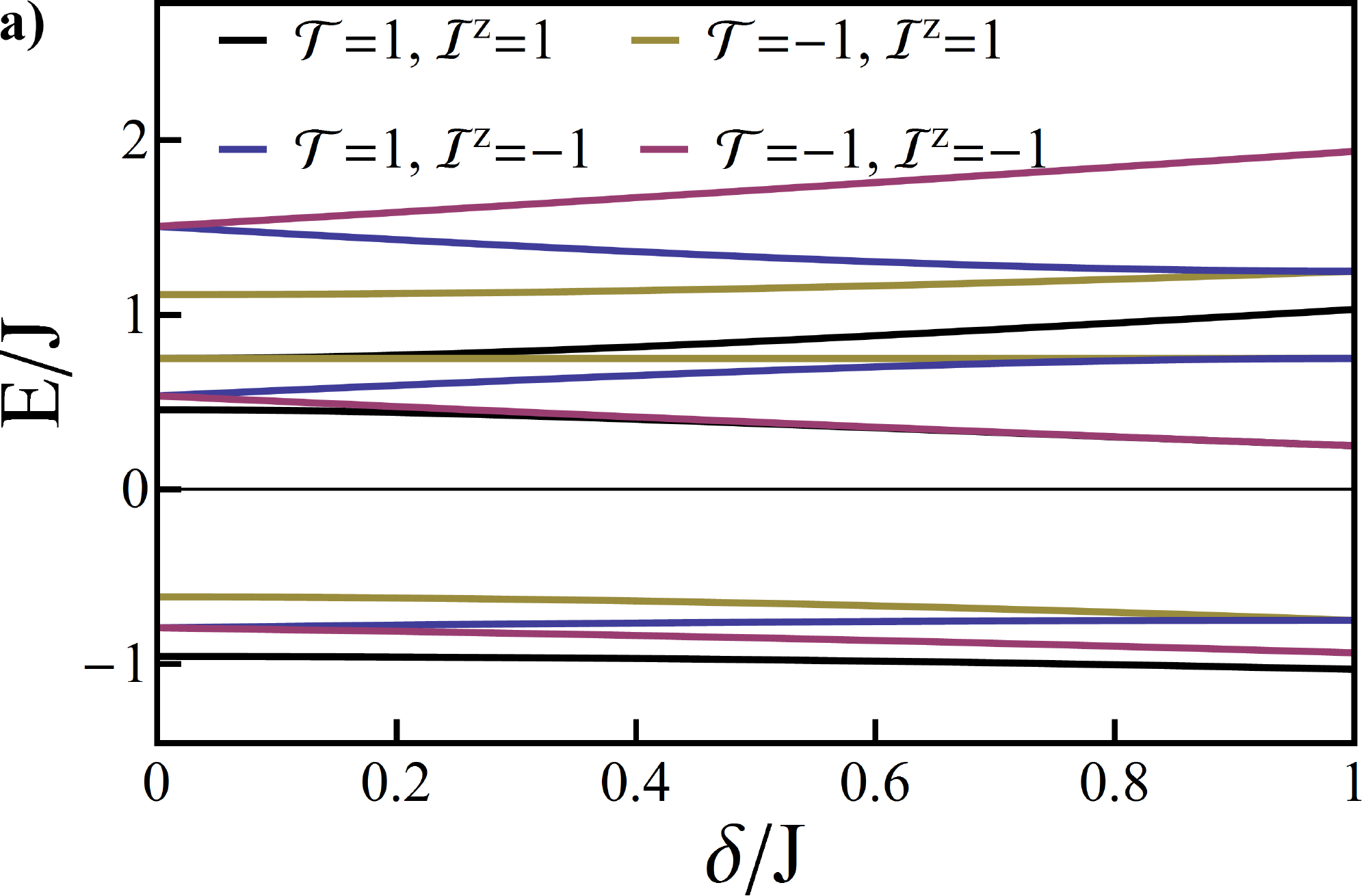}  \vspace*{10pt}\\	
\includegraphics[width=0.9\columnwidth]{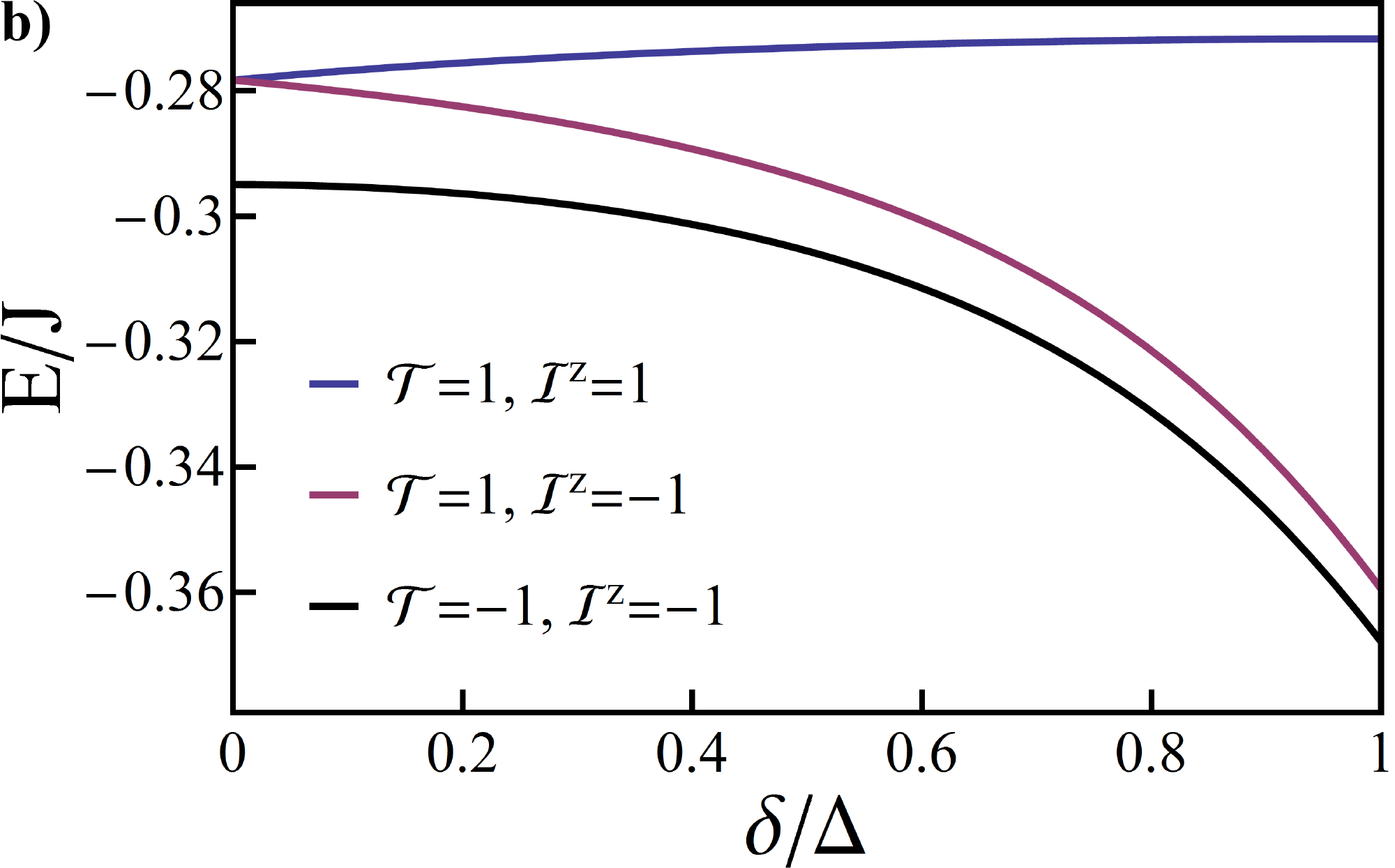}  	
\caption{Solution of the pseudo-angular momentum model of a complex with broken trigonal symmetry - due either to chemical modification or excited state localization. Panel (a) shows the full spectrum for states with $L^z=1$. Panel (b) shows only the $T_1$ substates, which are our primary concern. Here we take $\Delta=J/2$ and $\lambda=J/5$.}\label{fig:JT}
\end{center}
\end{figure}

We plot the spectrum of ${\cal L}^z=1$ states in Fig. \ref{fig:JT}a. In Fig \ref{fig:JT}b, we plot the same results, but only show the three lowest energy substates, I-III, which are of primary technological interest. One sees that the although states II and III are degenerate at $\delta=0$, a Jahn-Teller distortion rapidly lifts this degeneracy and for reasonable values of $\delta$ one finds that there is a much smaller energy gap between substates I and II than between II and III. This is what is observed experimentally \cite{R1,R19,R20} in a huge range of complexes (Table \ref{tab:expt}). We will see below that this splitting is the signature of the localization of the excitation to a single ligand.

Note, in particular, that substate I remains an admixture of $|T_\mathbb{1}\rangle$  with $|T_{z^2}\rangle$  and $|T_{x^2-y^2}\rangle$ and  has quantum numbers ${\cal T}={\cal I}^z={\cal L}^z=+1$. As none of the singlet states have these quantum numbers, cf. Table \ref{Tab:basis}, this state is forbidden from mixing with any of the singlet states in the model by conservation of ${\cal T}$, ${\cal I}^z$ and ${\cal L}^z$. Therefore substate I remains a pure triplet and is forbidden from decaying radiatively, irrespective of the strength of the SOC.

The radiative rate of the $m$th eigenstates of the full Hamiltonian, $|\psi_m\rangle$, is given by 
\begin{eqnarray}
k_R^m=\frac{4m_e e^4\alpha E_m^3}{3(4\pi\varepsilon_0)^2\hbar^3} \sum_{\footnotesize {
\begin{array}{c}
\beta\in\{x,y,z\} \\ n\in\{x,y,z\}
\end{array}}
}
\langle S_0|\mu_\beta|S_n\rangle\langle S_n|\psi_m\rangle
\end{eqnarray}
where $E_m$ is the excitation energy of the $m$th state, and $\alpha$ is the fine structure constant. Because of the underlying octahedral symmetry we take $\langle S_0|\mu_\beta|S_n\rangle$ to be independent of $n$ and further we assume that the zero field splitting is small compared to the $S_0\rightarrow T_1$ excitation energy, i.e., that $E_I\simeq E_{II}\simeq E_{III}$. It is also convenient to define 
\begin{eqnarray}
k_R^S=\frac{4m_e e^4\alpha E_m^3}{9(4\pi\varepsilon_0)^2\hbar^3} \sum_{\footnotesize {
\begin{array}{c}
\beta\in\{x,y,z\} \\ n\in\{x,y,z\}
\end{array}}
}
\langle S_0|\mu_\beta|S_n\rangle
\end{eqnarray}
this corresponds to the radiative decay rate for a pure singlet with an excitation energy equal to that of the $T_1$ manifold. 

\begin{figure}
\begin{center}
\includegraphics[width=0.9\columnwidth]{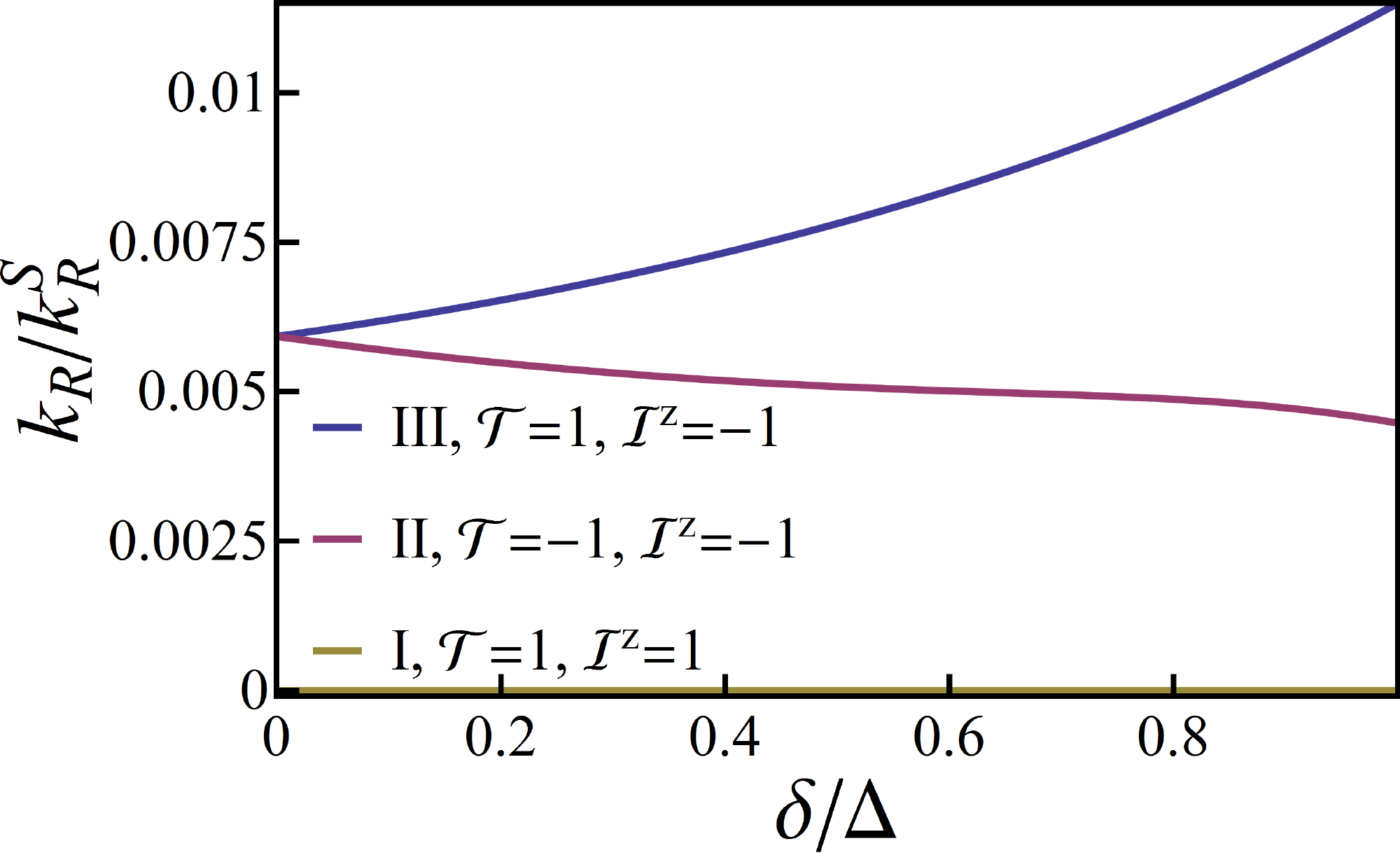}  
\caption{The radiative decay rates of the three substates of $T_1$. The conservation of $\cal T$,  ${\cal I}^z$, and ${\cal L}^z$ leads to the absence of radiative decay for state I. It can be seen that once the Jahn-Teller distortion becomes significant the radiative decay rate from state II is significantly smaller than that from state III, in good agreement with experiment (cf. Table \ref{tab:expt}). Here, as above, we take $\Delta=J/2$ and $\lambda=J/5$.}\label{fig:rates}
\end{center}
\end{figure}

We plot the radiative decay rate  in Fig. \ref{fig:rates}. State I is dark -- as expected from the conservation laws derived above. Furthermore, once the Jahn-Teller distortion becomes significant one finds that the radiative decay from state II is significantly slower than the radiative decay from state III. This is in precisely what is observed in experiments \cite{R1,R19,R20} on pseudo-octahedral $d^6$ complexes (cf. Table \ref{tab:expt}). 

It is straightforward to understand both the changes in energy and the radiative rates of states II and III. For $\delta>0$ ($\delta<0$ simple reverses these effects) the trigonal perturbation lowers the energy of (stabilises) states that are antibonding between the ${\cal I}^z=\pm1$ orbitals, e.g., $|S_x\rangle$ and $|T_x\rangle$, and raises the energy of (destabilises) those that are bonding between the ${\cal I}^z=\pm1$ orbitals, e.g., $|S_y\rangle$ and $|T_y\rangle$. It is clear from Table \ref{Tab:basis} that whereas $|S_x\rangle$ and $|T_y\rangle$ are even under time reversal $|T_x\rangle$ and $|S_y\rangle$ are odd. Thus, SOC mixes $|S_x\rangle$ with $|T_y\rangle$ and $|S_y\rangle$ with $|T_x\rangle$. Hence the trigonal distortion increases the energy difference between the triplet and singlet basis states that contribute to state II (i.e., $|S_x\rangle$ and $|T_y\rangle$ for $\delta>0$); whereas trigonal symmetry reduces the energy difference between the triplet and singlet basis states that contribute to state III ($|S_y\rangle$ and $|T_x\rangle$ for $\delta>0$). Thus the symmetry of the model dictates that $k_R^I<k_R^{II}<k_R^{III}$, as is observed experimentally \cite{R1,R19,R20}, see Table \ref{tab:expt}.

\begin{figure}
\begin{center}
\includegraphics[width=0.9\columnwidth]{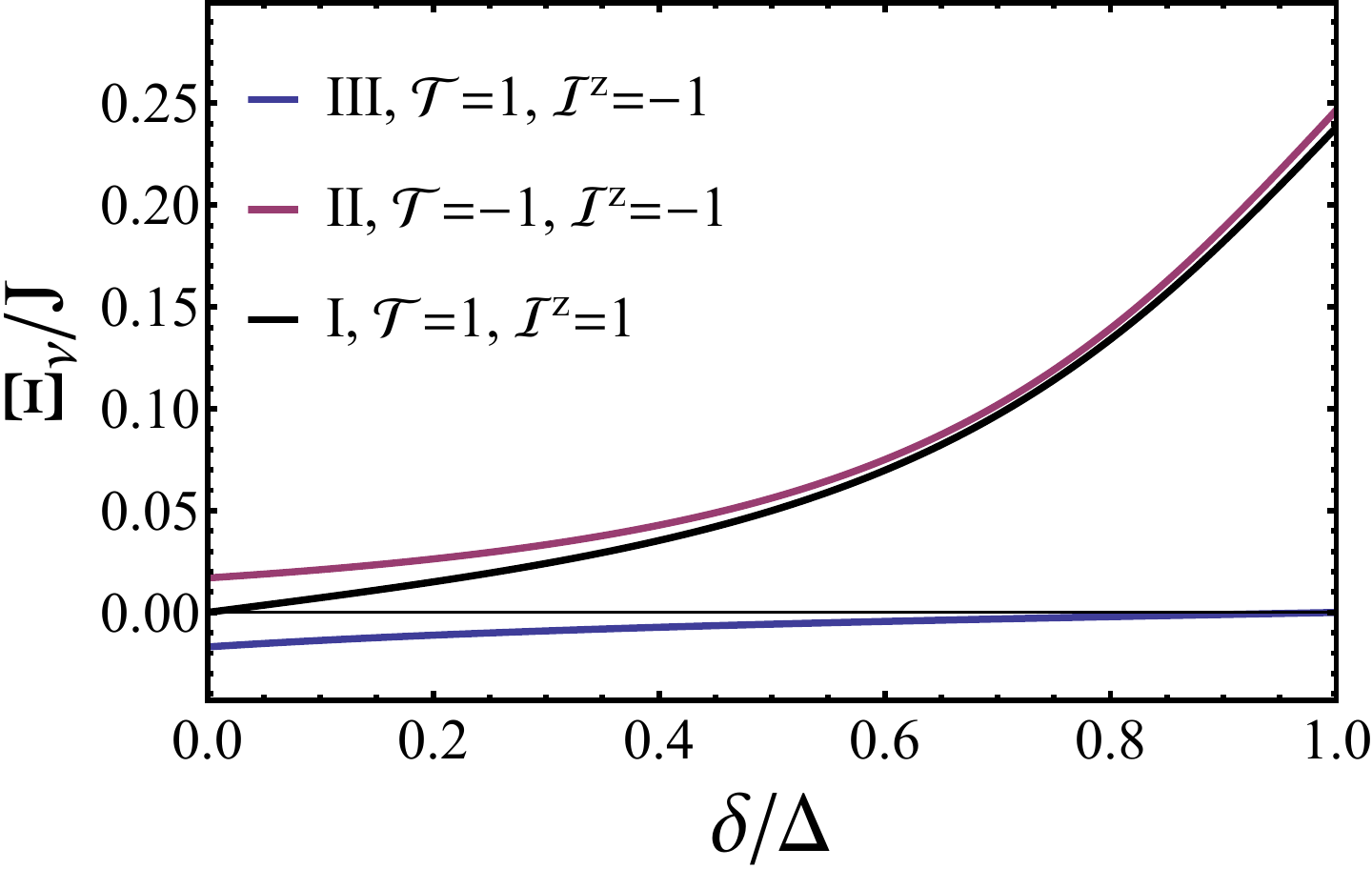}  
\caption{The degree of localization in the three substates of $T_1$, $\nu\in\{\textrm{I, II, III}\}$, where $\Xi_\nu=\sum_\sigma\left\langle\nu\left| a_{0\sigma}^\dagger a_{0\sigma} - \frac12\left( a_{1\sigma}^\dagger a_{1\sigma} + a_{2\sigma}^\dagger a_{2\sigma} \right) \right|\nu \right\rangle$. Here, as above, we take $\Delta=J/2$ and $\lambda=J/5$.} \label{fig:loc}
\end{center}
\end{figure}

Finally, we turn to the question of localization in the excited state. To measure this we define
\begin{eqnarray}
\Xi_\psi=\sum_\sigma\left\langle \psi\left|  a_{0\sigma}^\dagger a_{0\sigma} - \frac12\left( a_{1\sigma}^\dagger a_{1\sigma} + a_{2\sigma}^\dagger a_{2\sigma} \right) \right|\psi\right\rangle.
\end{eqnarray}
We plot $\Xi_\psi$ for the three substates of $T_1$ in Fig \ref{fig:loc}. The lowest energy excitation, I,  is completley delocalized for $\delta=0$ but rapidly  localizes for $\delta>0$. It is interesting to note that both  $\Xi_\textrm{II}$ and  $\Xi_\textrm{III}$ are non-zero for $\delta=0$.  However, for $\delta=0$ states II and III are degenerate and  $\Xi_\textrm{II}=-\Xi_\textrm{III}$, consistent with  trigonal symmetry. (A negative value of $\Xi_\psi$ indicates that the state disfavors occupation of site 0.) Nevertheless, for $\delta>0$, one observes a rapid increase in $\Xi_\textrm{II}$  whereas $\Xi_\textrm{III}$ grows only rather slowly.

It is therefore clear that the pseudo-angular momentum model predicts significant localization for values of $\delta$ compatible with the observed experimental results that $k_R^\textrm{I}<k_R^\textrm{II}<k_R^\textrm{III}$ and $E_\textrm{I,II}<E_\textrm{II,III}$, cf. Table \ref{tab:expt}. We therefore conclude that all of the complexes in Table \ref{tab:expt} show significant localization in their excited states. 

\section{Conclusions}

The pseudo-angular momentum model gives a natural explanation of the zero-field splitting observed in a wide range of pseudo-octahedral $d^6$ organometallic complexes. Furthermore, the conservation laws, and hence selection rules, inherent in the model give a natural explanation of the relative radiative decay rates of the three sublevels of $T_1$. We stress that none of the results derived here rely on perturbation theory -- therefore these conclusions hold even when the departures from octahedral or trigonal symmetry are large. This immediately explains why the properties of the $T_1$ states are so similar in both homoleptic and heteroleptic complexes.  Furthermore, for parameters compatible with the experimentally measured energies and radiative rates of the substates of $T_1$, the pseudo-angular momentum model predicts that exciations I and II are strongly localised -- although III remians well delocalised. Thus we conclude that all of the complexes in Table \ref{tab:expt} show significant localization in their two lowest energy excited (sub)states. 

It is interesting to note that when the radiative rates of individual sublevels, $k_R^m$, have been measured, rather than excited state lifetimes, $\tau_m$, it is found that the relative rates are in good accord \cite{R1,R19}, cf. Table \ref{tab:expt}. This is consistent with the high photoluminscent quantum yields observed in these complexes. This suggests that the non-radiative decay rates of the individual sublevels are determined by similar conservation laws. Therefore, it would be interesting to investigate non-radiative decay rates in a suitable extension of the pseudo-angular momentum model. 

We note that the pseudo-angular momentum model described above can be naturally extended to understand the properties other molecules and complexes where the low-energy excited states correspond to transition between degenerate or approximately degenerate states.

\section*{Acknowledgments} 
I think Sam Greer, Anthony Jacko and Ross McKenzie for helpful conversations. The work was supported by the Australian Research Council under grant FT130100161.

\section{Appendix: Estimation of parameters}

While we will not make a detailed parameterization of this model -- an idea of the relevant parameter ranges can be obtained from previous experiments and density functional calculations. Ir(ppy)$_3$ has been particularly widely studied and so is an ideal material to compare with. $\theta\sim\pi/4$ as the HOMOs are found to have about 50~\% metallic weight \cite{R8,R11,Hay,R10}. Nozaki \cite{R8} found that for Ir $\lambda_m=550$~meV and $\sin^2\theta=0.4$, yielding $\lambda=220$~meV. Smith {\it et al.} \cite{R11} considered the C$_3$ $S_0$ geometry and found from ground state calculations show that the gap between the HOMO and HOMO-1 is 140 meV (LUMO and LUMO+1 is 90 meV), which may be taken as an estimate of $\Delta$ ($\Gamma$). However several authors Nozaki \cite{R8,R17,JackoJCP} have noted that the values of $\Delta$ and $\Gamma$ are difficult to calculate from first principles -- therefore these numbers should be treated with some caution and are likely to be underestimates as interactions significantly increase the effective values of $\Delta$ and $\Gamma$. For ppy it has been estimated \cite{JackoJCP} based on the absorption spectra, emission spectra, and emission lifetimes \cite{Maestri} that $J_\pi\sim2$~eV; and for an isolated Ir ion $\lambda_d\sim0.43$~eV \cite{Vugman}.  Taking $\theta\sim\pi/4$ yields $J\sim1.4$~eV and $\lambda\sim300$~meV. For concreteness we  take $\lambda=J/5$ and $\Delta=J/2$ in the main text. However, our results are insensitive to the values of these parameters -- to demonstrate this we explore a range of other parameters in the sup. info. $\delta$ and $\gamma$ are not straightforward to estimate from previous work and will be left as free parameters, however as the C$_3$ symmetry remains evident even in the $T_1$ geometry of the excited this suggest that $\delta$ ($\gamma$) is not significantly larger than $\Delta$ ($\Gamma$).


\begin{thebibliography}{100}
\bibitem{footKM} Note that, in contrast, Kober and Meyer \cite{R14} take $\Gamma<0$, which means that their results, from conventional ligand field theory, are very different from ours.
\end{thebibliography}
\end{document}